\documentclass[12pt]{article}

\usepackage{graphicx}
\usepackage{epsfig}
\usepackage{amsfonts}

\newcommand{\la}{\lambda}
\newcommand{\ka}{\kappa}
\newcommand{\al}{\alpha}

\newcommand{\ta}{\theta}
\newcommand{\f}{\phi}
\newcommand{\vf}{\varphi}

\newcommand{\ee}{\end{equation}}
\newcommand{\eea}{\end{eqnarray}}
\newcommand{\be}{\begin{equation}}
\newcommand{\bea}{\begin{eqnarray}}

\newcommand{\pa}{\partial}

\newcommand{\vep}{\varepsilon}

\newcommand{\re}[1]{(\ref{#1})}
\newcommand{\R}{{\rm I \hspace{-0.52ex} R}}

\begin{document}
\begin{center}

{\Large Solitons and soliton--antisoliton pairs of a  
\\Goldstone model in $3+1$ dimensions}
\vspace{0.6cm}
\\
Vanush Paturyan$^{\ddagger}$,
 Eugen Radu$^{\dagger}$
and  D. H. Tchrakian$^{\dagger \star}$
\vspace{0.2cm}
\\
$^{\ddagger}${\small Department of
Computer Science, National University of Ireland Maynooth}
\\
$^{\dagger}${\small Department of
Mathematical Physics, National University of Ireland Maynooth,}
\\
$^{\star}${\small School of Theoretical Physics -- DIAS, 10 
Burlington
Road, Dublin 4, Ireland }
\end{center}
\begin{abstract}
We study finite energy topologically stable static solutions to a
global symmetry breaking model in $3+1$ dimensions described by an
isovector scalar field.
The basic features of two different types of configurations are 
studied, 
corresponding to axially symmetric multisolitons
with topological charge $n$, and unstable soliton--antisoliton pairs
with zero topological charge. 
\end{abstract}

\section{Introduction}
The familiar solitons in $3+1$ spacetime dimensions are the 
monopoles
\cite{mono} of the Yang-Mills--Higgs (YMH) model and the Skyrmions
\cite{skyrme}
of the $O(4)$ nonlinear sigma model. The first of these~\cite{mono} 
is a
solution of a gauged scalar (Higgs) field model, while the 
second~\cite{skyrme}
is not related to a gauge field. But this is not a very strict 
distinction,
since it is also possible to find solitons of the $SO(3)$ gauged 
$O(4)$
nonlinear sigma model \cite{gaugedo4}. These are all models that 
pertain to
physically rather different contexts but are nonetheless closely 
related
inasfar as they are all classical field theories whose static 
energies are
bounded from below by topological charges. On this rather technical 
level
therefore, one can ask whether there might be an ungauged Higgs 
analogue of the
the (ungauged) Skyrme soliton? We refer to such a symmetry--breaking 
field
theory, as a Goldstone model in $3+1$ dimensions.

It is known on the other hand that such models do exist \cite{gold-d} 
not just
in $3+1$ but in all dimensions. These are the gauge decoupled 
versions of the
$SO(D)$ gauged Higgs models \cite{sodhiggs} in $D+1$ spacetime 
dimensions. That
such models should support solitons follows from the simple fact that 
certain
truncated versions of these have such solutions in closed form 
\cite{gold-sd}.
The solitons of these models have the salient feature that the 
asymptotic
behaviour of their solitons feature the same properties as gauged 
Higgs
models, and hence afford a simple background for the study of Dirac
equations~\cite{dirac} in all dimensions.

While the existence of solitons to generic such Goldstone models were 
known
for sometime, a concrete and detailed construction of these has not 
been
presented in the literature to date. This is what we propose to do in 
this
paper, and since $3+1$ is physically the most relevant dimension, we 
have
chosen this for our example. The solitons we construct are in a 
sense
alternatives to the usual Skyrmions, though we have not pushed this 
analogy
here. Our strategy here is instead to expose the generic properties 
of such
solitons. To this end, we construct the topological charge-$1$ 
spherically
symmetric soliton, the axially symmetric winding number $n$ 
multisolitons (MS)
and examine the possiblity of the existence of bound states. We also
construct an (axially symmetric) topological charge-$0$ 
soliton-antisoliton
(SAS) pair, to highlight the analogy of the model studied with the 
usual
YMH model. 

In Section {\bf 2} we define the flat space energy density 
functional
of the model and the topological charge density presenting its lower 
bound.
In Section {\bf 3} we present the charge-$1$ solitons, the 
charge-$n$
MS, and the charge-$0$ SAS pairs, in
successive Subsections respectively, and in Section {\bf 4}
we summarise our results.

\section{The model and the topological charge}
The symmetry breaking model in 3 spatial dimensions, to which we 
refer as a
Goldstone model, is described by a scalar isovector field $\f^a$, 
$a=1,2,3$.
There is such a hierarchy of models \cite{gold-d} that arise from 
the
gauge decoupled limit of the three dimensional $SO(3)$ gauged Higgs 
model
descended from the $p$-th member of the Yang-Mills (YM) hierarchy on
$\R_3\times S^{4p-3}$. Here we have chosen the simplest of these, 
namely
that descended from the $2$-nd member of the YM hierarchy. Using the 
notation
\[
\f_i^a=\pa_i\f^a\quad,\quad\f_{ij}^{ab}=\pa_{[i}\f^a\pa_{j]}\f^b
\quad,\quad\f_{ijk}^{abc}=\pa_{[i}\f^a\pa_{j}\f^b\pa_{k]}\f^c\,,
\]
with the brackets $[ij...]$ implying total antisymmetrisation, the 
static
energy density functional is
\be
\label{en2}
{\cal E}_{(p=2)}=\la_0\,\left(\eta^2-|\f^a|^2\right)^4+\la_1\,
\left(\eta^2-|\f^b|^2\right)^2\,\left|\f_i^a\right|^2+
\la_2\left|\f_{ij}^{ab}\right|^2\, ,
\ee
which implies a total mass
$M=1/(4\pi)\int {\cal E} dV.$

All the dimensionless constants $\la_0$, $\la_1$ and $\la_2$ must be 
positive
if the topological lower bound to be introduced below is to be 
valid.
Moreover, any of these constants can also vanish, provided that the 
absence
of the corresponding term in \re{en} does not violate the Derrick 
scaling
requirement. That soliton solutions to this model exist is obvious 
since
for particular choices of these dimensionless constants, explicit 
solution
\cite{gold-sd} are known. Pushing our freedom of choosing the 
numerical values
of $\la_0$, $\la_1$ and $\la_2$ further, we can add any other 
positive
definite term to \re{en2} multiplying a new dimension dimensionless
coefficient, as long as the scaling properties remain satisfied. In 
$3$
spatial dimensions, there is one such possible kinetic term which for 
which
a canonical momentum field exists, and that is the {\it sextic}
term~\footnote{The correspoding {\it sextic} term in the Skyrme model 
was
considered in \cite{piette} in some detail.}. Thus the
most general model we can consider is the following extension of 
\re{en2}
\be
\label{en}
{\cal E}=\la\,V\left(\eta,|\f^a|\right)+\tau\,
\left[\left(\eta^2-|\f^b|^2\right)^2\,\left|\f_i^a\right|^2+
\frac{1}{4}\,\left|\f_{ij}^{ab}\right|^2\right]+\frac{\ka^4}{36}\,
\left|\f_{ijk}^{abc}\right|^2\,,
\ee
$\tau$ and $\la$ being dimensionless constants, $\ka$ is with 
dimension of
length, and $V\left(\eta,|\f^a|\right)$ a generic symmetry breaking 
potential.

It is perhaps in order to point out that \re{en} is an {\it ad hoc} 
model,
rather than a dimensionally descended model like \re{en2}. Indeed, a
{\it sextic} term does appear in the next one to \re{en2}
\be
\label{en3}
{\cal E}_{(p=3)}=\la_0\,\left(\eta^2-|\f^a|^2\right)^6+\la_1\,
\left(\eta^2-|\f^b|^2\right)^4\,\left|\f_i^a\right|^2+
\la_2\,\left(\eta^2-|\f^b|^2\right)^2\left|\f_{ij}^{ab}\right|^2+
\la_3\,\left|\f_{ijk}^{abc}\right|^2\,,
\ee
descended from $p=3$ YM, with which we are not concerned. We will 
restrict
our attention to \re{en}, and mainly to the particular case where 
$\la=\ka=0$,
which like the Skyrme model captures the main qualitative features of 
the
soliton.

The model \re{en} has certain remarkable similarities to the Skyrme
\cite{skyrme} model, and other properties that differ fundamentally. 
On the
similarity side, there is the obvious shared feature of the scaling 
of
the distinct terms in both three dimensional models. Also, for the 
important
special case with $\la=\ka=0$, the Bogomol'nyi equations are 
overdetermined
like in the Skyrme model and there exist no solutions saturating the
Bogomol'nyi bound.

On the contrasting side, the order parameter field $\f^a$ here is a 
relic
of a Higgs field and has the same dimensions ($L^{-1}$) as a 
connection,
and the finite enrgy conditions require the symmetry breaking 
boundary
condition
\be
\label{bc}
\lim_{r\to\infty}|\f^a|=\eta\,.
\ee
For the (more standard) case of multisolitons centred at the origin, 
the
boundary condition there is
\be
\label{bc1}
\lim_{r\to 0}|\f^a|=0\,.
\ee
For the unit charge spherically symmetric soliton, when the system is 
described
by a single function $h(r)$, the conditions \re{bc1} and \re{bc} 
(see
\re{asym} below) result in the {\it monopole} like asymptotics of our 
solitons,
which are qualitatively different from the {\it instanton} like 
asymptotics
of the Skyrmions~\footnote{The values of the scalar Higgs function
of a unit charge {\it monopole} on the boundaries $[r=0,r=\infty]$ 
are
$[h(0)=0,h(\infty)=1]$. The values of the scalar function $w(r)$ of 
the
spherically symmetric Yang--Mills (YM) {\it instanton} on the 
boundaries
$[r=0,r=\infty]$ are $[w(0)=\pm 1,w(\infty)=\pm 1]$. The one 
dimensional
reduced action density of the scale invariant YM system in $4p$ 
dimensions
is proportional to the corresponding $O(2p+1)$ sigma model reduced
action in $2p$
dimensions, described by the function $f(r)\equiv\arccos w(r)$. Hence 
the
values of the scalar function $\cos f(r)$ of the unit charge sigma 
model
soliton on the boundaries $[r=0,r=\infty]$ are
$[\cos f(0)=\pm 1,\cos f(\infty)=\pm 1]$, like an {\it instanton}.}.

It is straightforward to show that \re{en} is bounded
from below by the density
\bea
\varrho&=&\frac{1}{4\pi}\vep_{ijk}\vep^{abc}\,
\left(\eta^2-|\f^d|^2\right)\,\f_i^a\,\f_j^b\,\f_k^c\nonumber\\
&=&\frac{1}{4\pi}\vep_{ijk}\vep^{abc}\,\pa_i\,
\left[\left(\eta^2-\frac35|\f^d|^2\right)
\,\f^a\,\f_j^b\,\f_k^c\right]\label{totdiv}
\eea
whose volume integral is the topolgical charge, which is just the
{\it winding number}.

Models like \re{en2}, \re{en} and \re{en3} support {\it global 
monopoles}
in the sense that their topological charges are the winding numbers 
of the
scalar (Higgs) field on the 2-sphere at infinity. These solitons 
however
have finite energy, unlike the usual {\it global monopoles}. This is 
due
to two reasons. Firsly we have included the {\it quartic} kinetic 
term to
satisfy the required scaling, and secondly, we have employed a non
standard {\it quadratic} kinetic term which decays fast enough to
satify finite energy requirements. These two features are guaranteed
by the fact that these models are dimensionally descended from 
higher
dimensional Yang--Mills models, the latter being endowed with the
corresponding topological properties, as explained in \cite{gold-d} 
and
references therein.

While we are exclusively concerned with the classical properties of 
the
model \re{en} here, it is nevertheless interesting to comment on its
possible quantum aspects. It is believed that the Skyrme model is a
reliable approximate theory of the Nucleons~\cite{ANW} and our model 
shares
many similarities with the latter. The main difference between the 
models is
that while the Skyrme field is a constrained field and hence the 
procedure
of quantisation must take account of the constraint, the order 
parameter
field in our model is unconstrained. On the other hand, the {\it 
quadratic}
kinetic term in \re{en} is quite unconventional, featuring the field
dependent factor $\left(\eta^2-|\f^b|^2\right)^2$. The effect of the
latter is to prevent the definition
of a propagator, this difficulty of quantisation replacing the
{\it constraint} problem in the Skyrme model. But it is also well 
known that
the most efficient practical method of quantisating Skyrme theory is 
that
of {\it collective coordinate} quantisation, employed in \cite{ANW}. 
This
method applies is equally well to the system \re{en},
subject to making the essential change of field coordinates
\[
\f^a\rightarrow\Phi=\f^a\,\tau_a\ \ ,
\]
$\tau_a$ being the three Pauli matrices.

\section{The solitons}
Our aim in this Section is to demonstrate the close similarities of 
the
solitonic solutions in this model, with the various monopole (and 
dipole)
solutions of the YMH model. 
The Section is divided into three Subsections. The first involves 
the
charge-$1$ soliton analogous to the 't~Hooft--Polyakov 
monopole~\cite{mono}.
The second is concerned with axially symmetric solutions satisfying 
standard
boundary conditions, with arbitrary winding number $n$, namely the
multisolitons (MS) of this model. A question of interest raised in 
this
Subsection is that of the mutual attraction or repulsion two 
$1$-solitons.
In the last Subsection we impose those boundary conditions on the 
axially
symmetric fields, which result in zero charge (unstable) solutions 
representing
soliton--antisoliton (SAS) pairs situated on the symmetry axis. 
These
boundary conditions are those employed in \cite{KK,R} for the 
corresponding
$SO(3)$ YMH model.
\subsection{Charge-$1$ soliton: spherically symmetric}
Subjecting \re{en} to spherical symmetry via the Ansatz
\be
\label{sph}
\f^a=\eta\,h(r)\,\hat x^a\,,
\ee
and taking into account the factor $r^2$ in the volume element, the 
reduced
one dimensional energy functional, after some rescalings, is
\be
\label{eredfunc}
E=\la\,\eta^{-2}\,r^2\,V(\eta,h)+\tau\,\eta^6
\left[(1-h^2)^2\left(r^2\,h'^2+2h^2\right)
+\frac{h^2}{r^2}\left(2r^2\,h'^2+h^2\right)\right]+
(\eta\ka)^4\frac{h^4}{r^2}\,h'^2.
\ee 
\newpage
\setlength{\unitlength}{1cm}
\begin{picture}(6,8)
\centering
\put(1.6,0){\epsfig{file=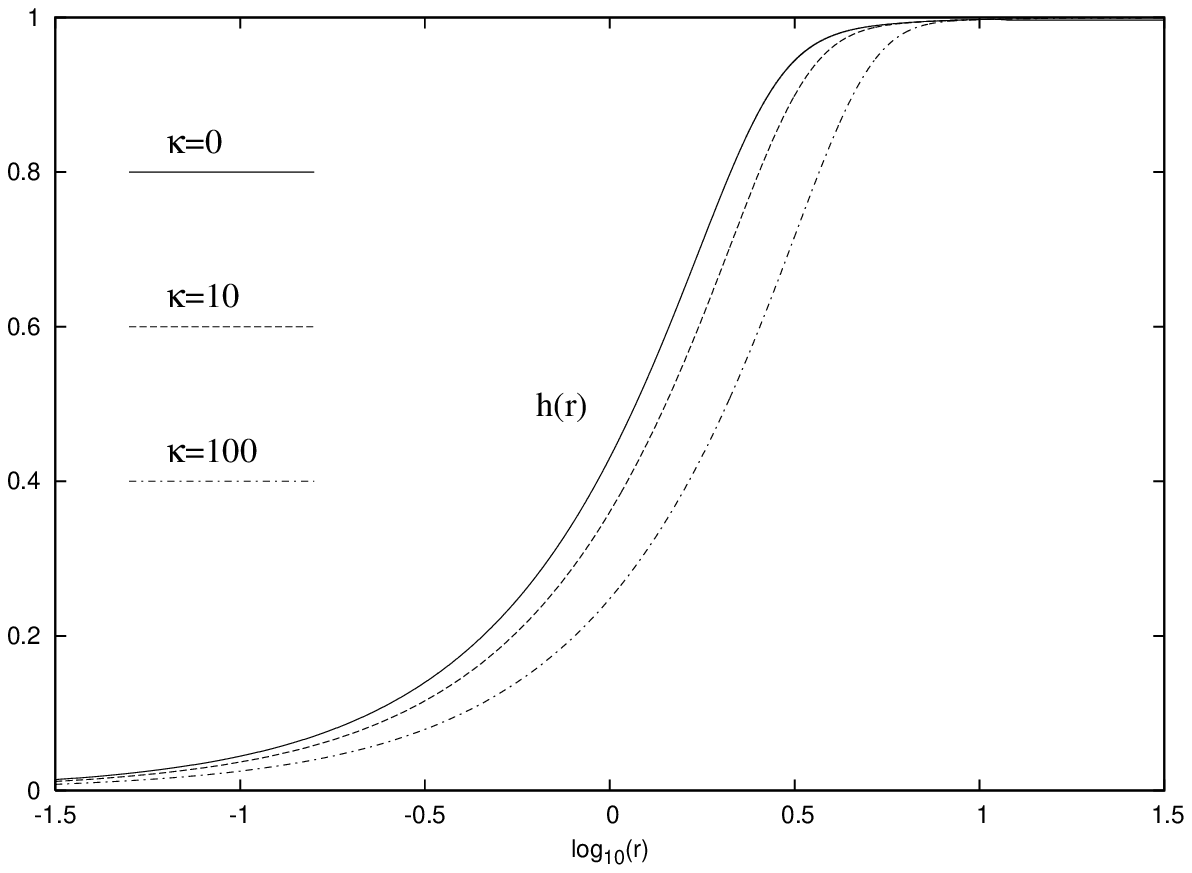,width=12cm}}
\end{picture}
\begin{center}
\end{center}
\begin{picture}(10,7.7)
\centering
\put(2.26,0){\epsfig{file=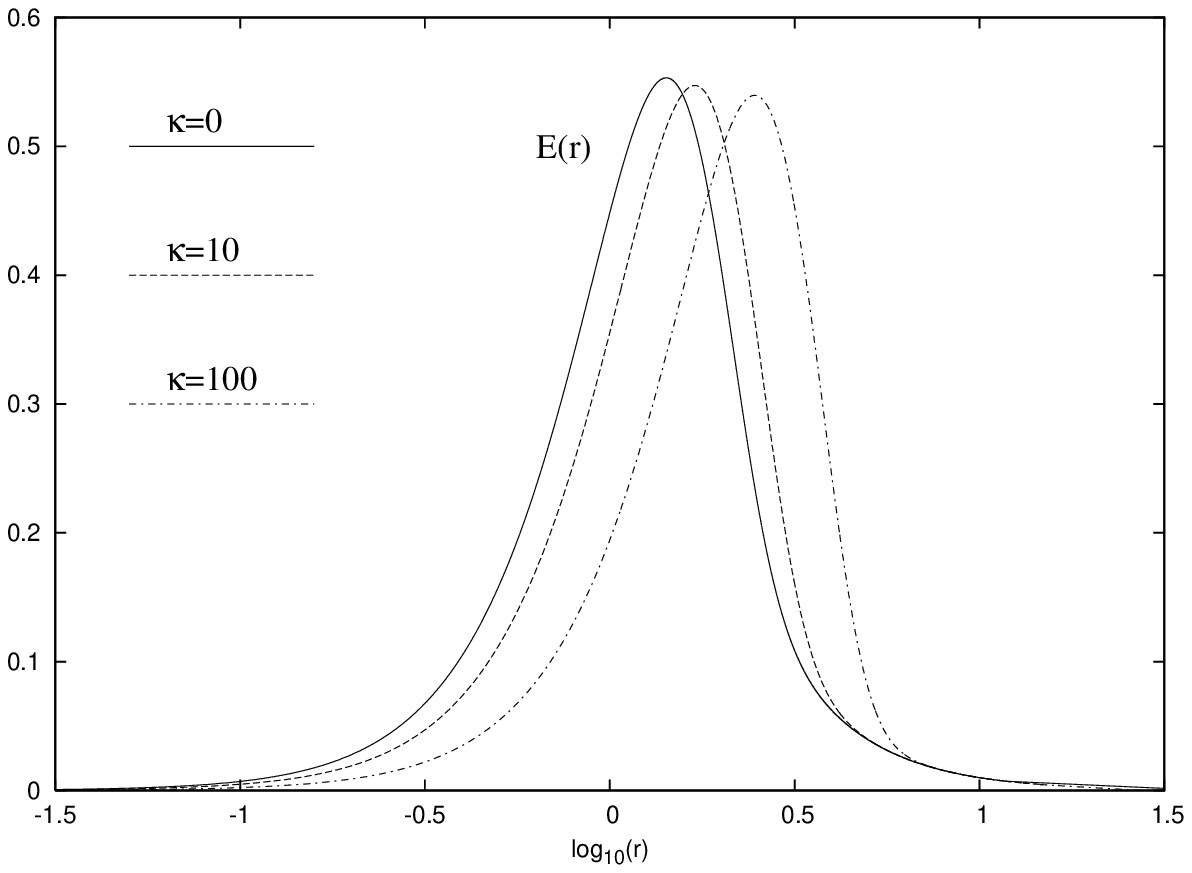,width=12cm}}
\end{picture}
\\
{\small {\bf Figure 1.}   The profiles of the function  $h(r)$ and 
the mass-energy  
$E(r)$ of typical spherically symmetric solutions
are shown for several values of $\kappa$.}
\\
\\
Mostly, we will retain only the terms multiplying $\tau$. In that 
case
the lower bound on the integral of \re{eredfunc} is
\be
\label{toplb}
Q=\int\,\varrho\,d^3x=
6\int_0^{\infty}\frac{d}{dr}\left(\frac{h^3}{3}-\frac{h^5}{5}\right)\,dr\,.
\ee
Substituting into the limits of the definite integral \re{toplb}\ ,
the asymptotic values
\be
\label{asym}
\lim_{r\to 0}h(r)=0\quad,\quad\lim_{r\to\infty}h(r)=1
\ee
following from \re{bc1} and \re{bc}, one finds
\[
Q=\frac45
\]
which was verified numerically.
The solutions of the  field equation  can be constructed 
numerically.
We follow the usual approach and, by using a standard ordinary
differential equation solver, we evaluate the initial condition 
\[
h=br-\frac{b^3}{5(b^2+2)}r^3+O(r^5)
\]
 at
$r=10^{-6}$ for global tolerance $10^{-14}$, adjusting for fixed 
shooting
parameter and integrating  towards  $r\to\infty$.
The behaviour of finite energy solutions as $r \to \infty$ is
 \[
h \sim 1+ce^{-2r}-\frac{1}{4r^2}-\frac{15}{32r^4} +O(1/r^6),
\]
where $c$ is a free parameter.
For all considered cases, solutions with the correct asymptotics 
occurs 
only when the first derivative of the scalar function $h(r)$ 
evaluated
at the origin, $h'(0)= b$,  takes on a certain value. For example
$b \sim 0.443613$ for a model without a sextic term ($\kappa=0$), 
while
$b \sim 0.367479$ for $\kappa=10$ 
(the symmetry breaking potential is vanishing in both cases).

The profiles of typical solutions are presented in Figure 1 
for several values of the parameter $\kappa$ and no symmetry 
breaking
potential. The energy functional,
as given by (\ref{eredfunc}) is also exhibited.
No multinode radial solutions were found, although we have no  
analytical argument for their absence.
However, our preliminary numerical results indicate that, 
similar to the case of  monopoles and sphalerons,
the gravitating  Goldstone model also presents radial excitations
 with an arbitrary number of nodes of the function $h(r)$.
A study of these solutions will be presented elsewhere.

\subsection{Charge-$n$ multisoliton: axially symmetric MS}
The axially symmetric Ansatz for the scalar field
\[
\f^a=(\f^{\al},\f^3)
\]
is
\be
\label{ax}
\f^{\al}=\eta\,\vf_1(\rho,z)\,\,n^{\al}\quad,\quad\f^3=\eta\,\vf_2(\rho,z)\,,
\ee
where $\rho^2=|x_{\al}|^2=x_1^2+x_2^2$, $z=x_3$, denoting 
$x_i=(x_{\al},x_3)$,
and $n^{\al}$ is the unit vector
\be
\label{n}
n^{\al}=(\cos n\,\f\,,\,\sin n\,\f)
\ee
with azimuthal winding $n$.

We denote the two functions $(\vf_1,\vf_2)\equiv\vf_A$ by labeling 
$\vf^A$
with $A=1,2$. Subjecting \re{en} to \re{ax}, the $\la=0$ system 
reduces to
\bea
E&=&2\pi\,\rho\,\Bigg\{
\tau_1^2\eta^6(1-|\vf_B|^2)^2\left[\left(|\pa_{\rho}\vf_A|^2+
|\pa_z\vf_A|^2\right)+\left(\frac{n\vf_1}{\rho}\right)^2\right]\nonumber\\
&&+\tau_2^2\eta^4\left[\left(\vep_{AB}\,\pa_{\rho}\vf_A\,\pa_z\vf_B\right)^2+
\left(\frac{n\vf_1}{\rho}\right)^2\left(|\pa_{\rho}\vf_A|^2+
|\pa_z\vf_A|^2\right)\right]\nonumber\\
&&+\ka^4\eta^6\left(\frac{n\vf_1}{\rho}\right)^2
\left(\vep_{AB}\,\pa_{\rho}\vf_A\,\pa_z\vf_B\right)^2\Bigg\}\,,\label{Erz}
\eea
which in terms of the more useful variables $(r,\theta)$ is
\bea
\label{energy}
E&=&4\pi\,\eta^6\,\tau_1^2\,\sin\theta\,\Bigg\{
(1-|\vf_B|^2)^2\left[\left(r^2|\pa_r\vf_A|^2+
|\pa_{\theta}\vf_A|^2\right)+\frac{n^2\vf_1^2}{\sin^2\theta}\right]
\nonumber\\
&&+\left(\frac{\tau_2}{\tau_1\eta}\right)^2
\left[\left(\vep_{AB}\,\pa_r\vf_A\,\pa_{\theta}\vf_B\right)^2+
\left(\frac{n\vf_1}{r\sin\ta}\right)^2\left(r^2|\pa_r\vf_A|^2+
|\pa_{\theta}\vf_A|^2\right)\right]\nonumber\\
&&+\left(\frac{\ka^2}{\tau_1}\right)^2
\left(\frac{n\vf_1}{r\sin\ta}\right)^2
\left(\vep_{AB}\,\pa_r\vf_A\,\pa_{\ta}\vf_B\right)^2
\Bigg\}\,,\label{Ert}
\eea
and rescaling $r$ as
\[
\tau\,r\equiv\left(\frac{\tau_2}{\tau_1\eta}\right)\,r\rightarrow\,r
\]
removes the coupling constant infront of the {\it quartic} term.

For the multisoliton (MS) solutions at hand,
the boundary values of the functions $\vf_A$ in the $r\gg 1$ region 
are
\be
\label{bcr>MS}
\lim_{r\to\infty}\vf_1(r,\ta)=\sin \ta\quad,
\quad\lim_{r\to\infty}\vf_1(r,\ta)=\cos \ta\,,
\ee
while at the origin we find
\be
\label{ta0MS}
\varphi_1|_{r=0}=\varphi_2|_{r=0}=0.
\ee
Since our imposition of axial symmetry requires also $z\to -z$ 
reflection
symmetry, the actual (numerical) integration need be performed only 
over
the range $0\le\ta\le\frac{\pi}{2}$.
The field equations have been solved by imposing the
following  boundary conditions along the axes
\begin{eqnarray}
\label{tapi2MS}
\varphi_1|_{\theta=0}=\partial_{\theta}\varphi_2|_{\theta=0}=0,
~~\partial_{\theta}\varphi_1|_{\theta=\pi/2}=\varphi_2|_{\theta=\pi}=0.
\end{eqnarray}
We solve numerically the
set of two coupled non-linear elliptic partial differential
equations arising from the
variation of the functional \re{Ert}, subject to the above boundary 
conditions,
employing a compactified radial coordinate $x=r/(1+r)$. 
To obtain  axially symmetric solutions, 
we start with the $n=1$ solution discussed above as initial guess
(corresponding to $\vf_1=h(r)\sin\theta $, $\vf_2=h(r)\cos\theta $) 
and
increase the value of $n$ slowly.
The iterations converge, and repeating the procedure one obtains
in this way solutions for arbitrary $n$.
The physical values of $n$ are integers.
The typical numerical error for the functions is estimated to be 
lower than $10^{-3}$. 
The numerical calculations for $n>1$ were performed with the software 
package 
CADSOL/FIDISOL, based on the Newton-Raphson method \cite{FIDISOL}.
In Figure 2 we show the local mass-energy as given by (\ref{energy}) 
 of the $\ka=0$  $n=2$ MS 
solution
as function of the coordinates $z=r \cos\theta$ and 
$\rho=r\sin\theta$.
In Figure 3
the profiles scalar functions $\varphi_1$ and $\varphi_2$ of the same 
solution
are shown for several angles  as a function of the radial coordinate 
$r$.
\newpage
\setlength{\unitlength}{1cm}
\begin{picture}(6,8)
\centering
\put(1.6,0){\epsfig{file=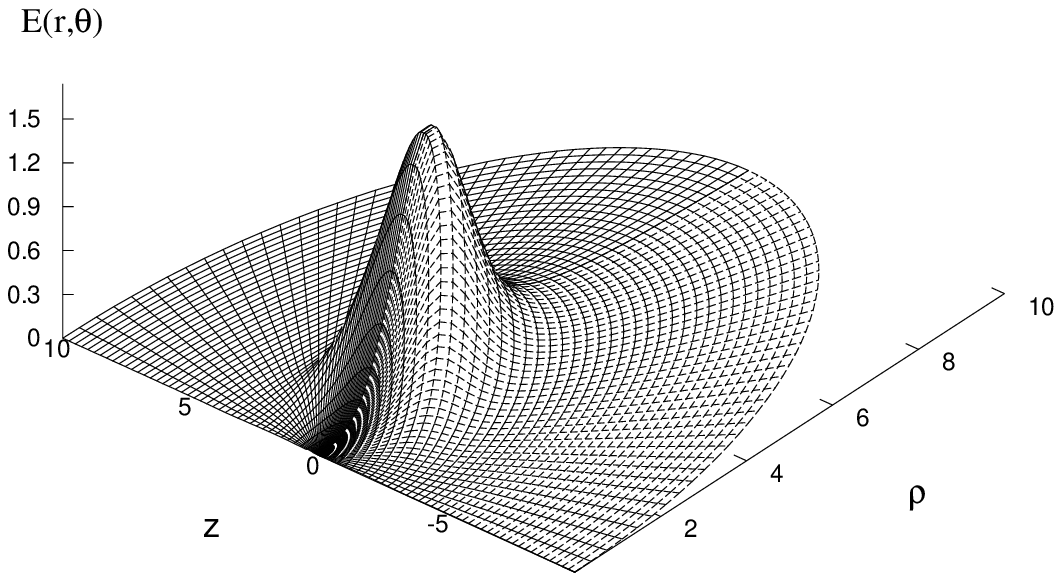,width=12cm}}
\end{picture}
\\
\\
{\small {\bf Figure 2.}   A three-dimensional plot of the mass-energy  
$E(r,\theta)$ of a $n=2,~\kappa=0$ axially symmetric MS solution. }
\\
\\
The analysis was carried out in the first place setting the constant 
$\ka=0$
and for several values of $n$,
which captures the main qualitative properties of the MS.
The maximum of the mass-energy density (\ref{energy}) moves outwards
with increasing  $n$. However, for $n>3$, the numerical errors start 
to
increase, and for some $n_{max}$ the numerical
iterations fail to converge. The problem resides in the behaviour of 
the 
scalar function $\varphi_2$, which for large $n$, 
tends to develop a discontinuity for some value of the
radial coordinate.

Because of the
close analogy between our model and the Skyrme model, it is 
worthwhile
checking one of the remakable properties of axially symmetric 
multi-Skyrmions.
The property in question is that up to vorticity ($\equiv$ baryon 
number)
$n=4$ the energy of the multi-Skyrmion is smaller than that of $n$ 
infinitely
separated $1$-Skyrmions, i.e. that the MS can be regarded as a bound 
state
\cite{Kopel}.

We have here checked that starting from $n=2$, and up to $n=5$, the 
energies of
the $n$-MSs of our model are greater than that of $n$ $1$-solitons. 
Moreover it
turns out that this deficit of binding energy increases with 
increasing $n$,
indicating that none of the MSs in this model can be regarded as 
bound states.
For example we have found $M(n=2)/(2M(n=1))-1=0.229$,
$M(n=3)/(3M(n=1))-1=0.381$
while $M(n=4)/(4M(n=1))-1=0.492$ (where $M(n=1)=1.188$). 

It is for this reason that we have introduced the 
sextic~\footnote{The presence
of higher order terms in the (covariant) derivatives of the Higgs 
field is
known to result in the mutual attraction of like charged (monopoles) 
solitons
\cite{OKT,GST}.} term in \re{Ert}, to check whether its presence may 
reverse
this trend and lead to MS bound states? However, we find that for
$0\leq\ka\leq 10$, MSs with
charges up to $3$ the deficit of binding energy persists and 
increases with
$n$, confirming that like charged solitons of this model are 
mutually
repulsive inspite of the 

\newpage
\setlength{\unitlength}{1cm}
\begin{picture}(6,8)
\centering
\put(1.6,0){\epsfig{file=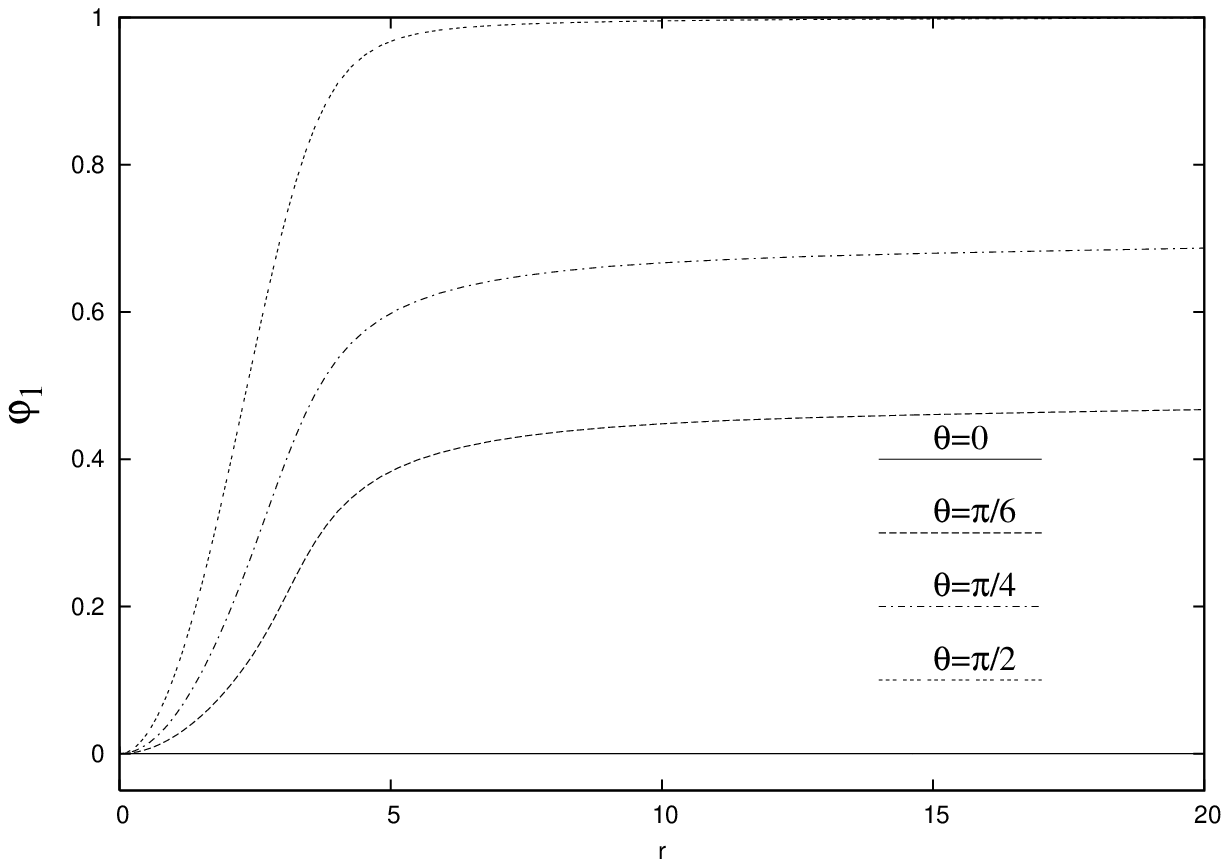,width=12cm}}
\end{picture}
\begin{center}
\end{center}
\begin{picture}(10,7.7)
\centering
\put(2.26,0){\epsfig{file=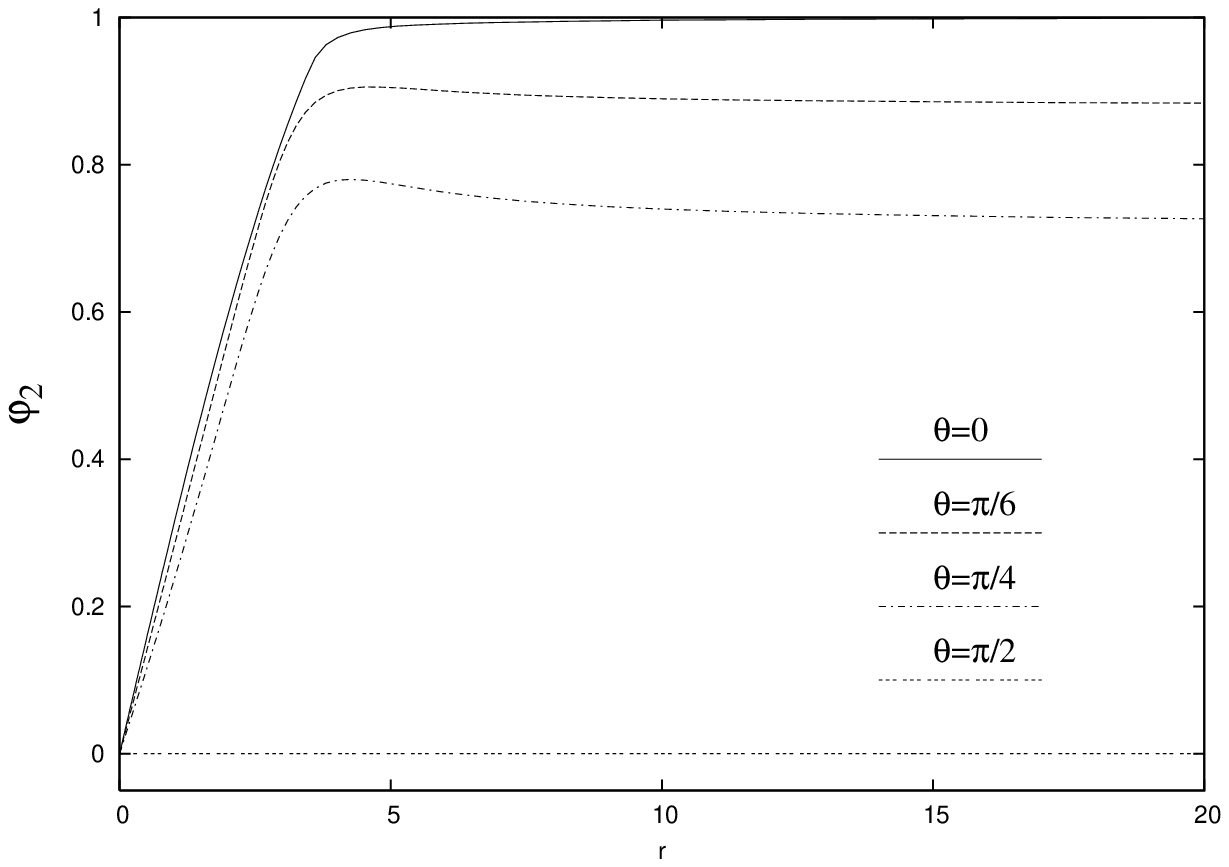,width=12cm}}
\end{picture}
\\
{\small {\bf Figure 3.}   The profiles of the scalar functions 
$\varphi_1$ and $\varphi_2$ are shown for
the  $n=2,~\kappa=0$ axially symmetric MS solution.}
\\
\\
model having descended from the gauge decoupling of
Higgs models~\cite{OKT,GST} supporting mutually attracting monopoles 
of like
charges.  

\subsection{Charge-$0$ soliton--antisoliton: axially symmetric SAS}
For simplicity we will restrict to winding number $1$ SAS
solutions, whence we set $n=1$ in the Ansatz \re{ax}.
\newpage
\setlength{\unitlength}{1cm}
\begin{picture}(6,8)
\centering
\put(1.6,0){\epsfig{file=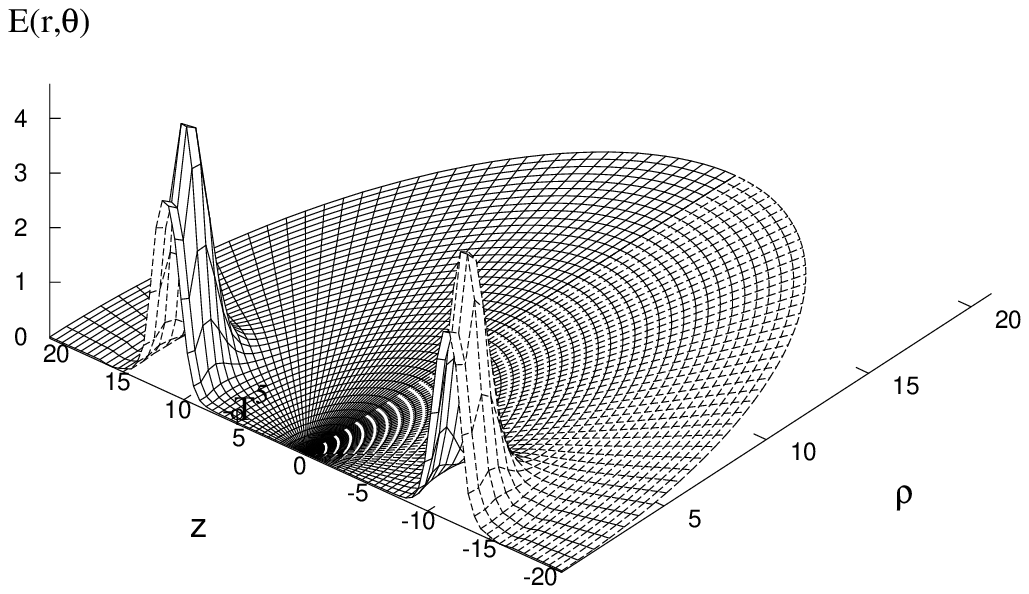,width=12cm}}
\end{picture}
\\
\\
{\small {\bf Figure 4.}    A three-dimensional plot of the 
mass-energy  
$E(r,\theta)$ of a $n=1,~\kappa=0$ axially symmetric SAS solution. }
\\
\\
The reduced two dimensional energy density functionals \re{Erz} and 
\re{Ert}
are unchanged for the SAS solutions. Also the boundary conditions  
\re{tapi2MS} arising from the requirements of analyticity on
the $z$-axis and at the origin, remain unchanged. What are different 
between
the MS and SAS solutions are the boundary conditions at $r\to\infty$ 
and at
$r=0$.

For the SAS solutions in the region $r\gg 1$, instead of \re{bcr>MS} 
we require
\be
\label{bcr>SA}
\lim_{r\to\infty}\vf_1(r,\ta)=\sin m\ta\quad,
\quad\lim_{r\to\infty}\vf_1(r,\ta)=\cos m\ta\,,
\ee
with $m\ge 2$ for SAS chains analogous to the monopole antimonopole 
chains
\cite{KKS}. But we are here only interested in SAS pairs, therefore 
we
restrict to $m=2$ in \re{bcr>SA}.

For the SAS in the region $r\ll 1$, instead of \re{ta0MS} we require
\be
\label{r0SA}
\vf_1|_{r=0}=0\quad,\quad\pa_{r}\vf_2|_{r=0}=0.
\ee 
The field equations have been solved by using the same methods 
employed
in the MS case. However, the SAS solutions exhibit a very different 
picture.
The energy density
$\epsilon = -T_{t}^t$ possesses maxima at $z=\pm d/2$ and a saddle
point at the origin, and presents the typical form exhibited in the
literature on MA solutions \cite{KK}.
The modulus of the scalar field 
$|\varphi|=\sqrt{\varphi_1^2+\varphi_2^2}$
possesses always two zeros at $\pm d/2$ on the $z-$symmetry axis.
In Figure 4 we plot the  mass-energy (\ref{energy}) and the
modulus of the scalar field of a typical $m=2$ solution as a function 
of the
coordinates $\rho, z$, for $\kappa=0$ ($i.e.$ no sextic term).
This solution has a mass $M=2.588$ which is smaller than that of two
($n=1,~m=1$) $1$-solitons (M=2.92), similar to the sphaleron 
describing a
monopole-antimonopole pair \cite{KK} .

\newpage
\setlength{\unitlength}{1cm}
\begin{picture}(6,8)
\centering
\put(1.6,0){\epsfig{file=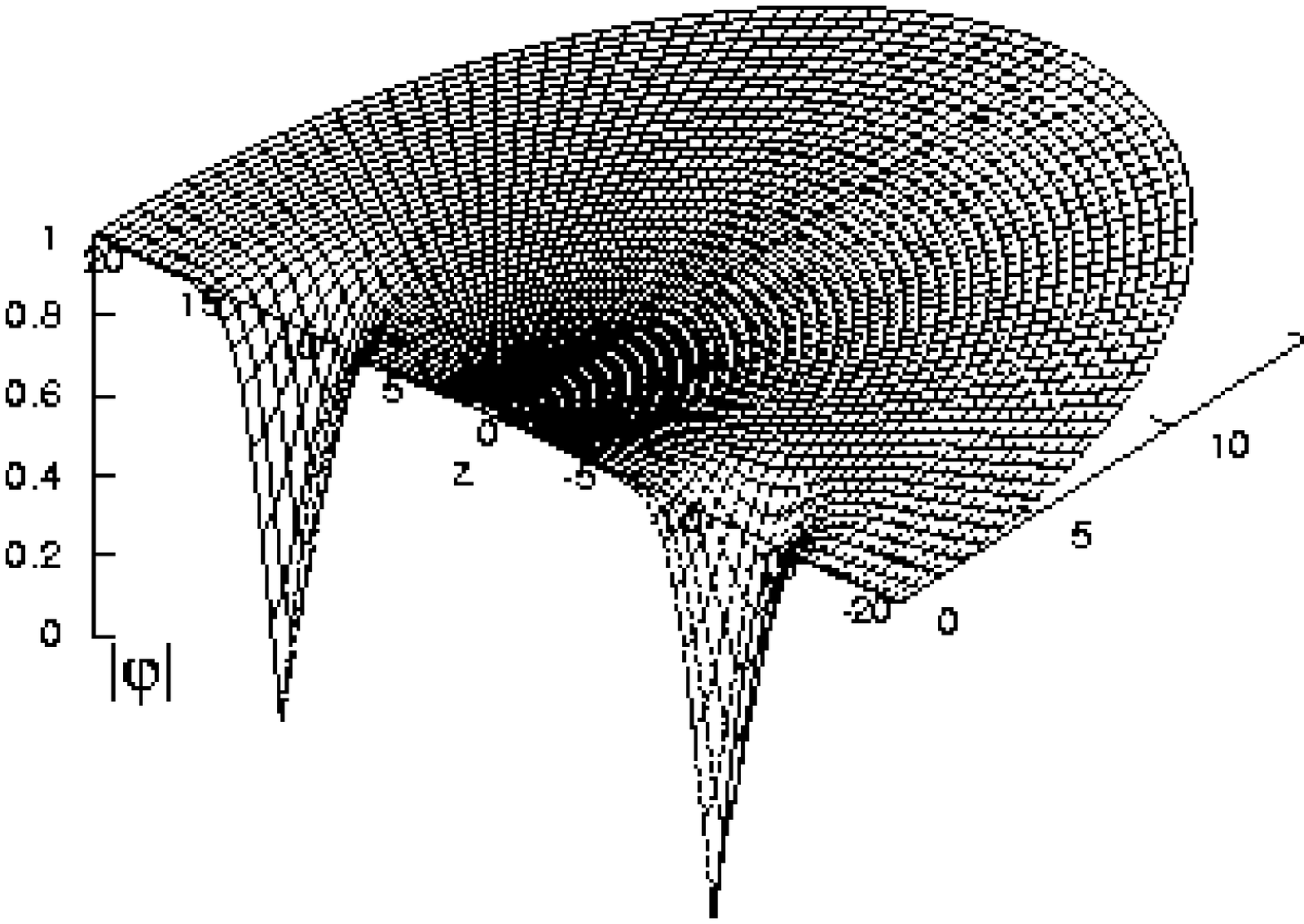,width=12cm}}
\end{picture}
\\
\\
{\small {\bf Figure 5.}    The modulus of the scalar field 
$|\varphi|=\sqrt{\varphi_1^2+\varphi_2^2}$ is shown 
for the $n=1,~\kappa=0$ axially symmetric SAS solution. }
%
\section{Summary}
We have studied the finite energy topologically stable static 
solutions to a
(global) symmetry breaking model in $3+1$ dimensions described by an
isovector scalar field. Such models can be constructed in arbitrary
$D+1$ dimensions since they are the gauge decoupled versions of Higgs 
models
in all dimensions. We have chosen here $D=3$ examples, since this is 
the
dimensionality of most physical interest, like the usual Skyrme 
model, but
very different from the latter in many essential respects.

Two classes of solutions are studied: Axially symmetric 
multisolitons
(MS) with topological charge $n$, and unstable soliton--antisoliton 
(SAS) pairs
with zero topological charge, both with finite energies. There are 
two
pertinent questions that arise here. In the case of the MS solutions, 
the
question is whether solitons of like charge attract or repel, and it 
was found
that they always repel, even when the model is augmented with a 
sextic kinetic
term. In the case of the SAS pairs, the question is whether they can 
support
a nonzero angular momentum \footnote{The correspoding rotating 
solutions in YMH model have 
been studied recently in \cite{Paturyan:2004ps} and 
\cite{Kleihaus:2005fs}.} ?
This task is deferred to some future work, and
presumably it will involve a stationary Q-ball like features.

As a scalar theory supporting soliton solutions in 3 space dimensions 
this
model is like the Skyrme model. Unlike the latter however this is a
symmetry breaking model, as result of which the boundary values of 
the
field are akin to that of a monopole rather than that of an instanton 
as
is the case for the Skyrme (nonlinear sigma) model. From the 
viewpoint of
physical properties, there is no question that it can be regarded as 
an
alternative for the Skyrme model which is known to give a good 
description
of Nucleons~\cite{ANW} at low energies. This can be seen from two 
clear
viewpoints: ({\it i}) the fact that
Skyrmions are capable of forming bound states~\cite{Kopel} 
describing
exotic states, while the MSs of our model display the opposite 
property, and
({\it ii}) because the Skyrmion can be gauged with the (Maxwell) 
$U(1)$
field~\cite{PT, Radu:2005jp} enabling the description of the 
electromagnetic properties
of the Nucleons, while the topological lower bound on the energy of 
our
MS is invalidated when the scalar field is gauged with $U(1)$. This 
is
because Higgs models, from the $p=2$ member of which~\cite{sodhiggs} 
the
present model is extracted, can be gauged only with $SO(D)$, or, be 
completely
gauge-decoupled as the model considered here. While it is true that
a $O(D+1)$ sigma model in $D$ dimensions can be gauged with all
$SO(N)$ with $N\le D$~\cite{PT} with its energy bounded from below by 
a
gauge invariant topological charge, gauging Higgs models 
\cite{sodhiggs}
with $SO(N)$ with $1<N<D$ causes the collapse of the topological 
lower bound
on the energy.

Technically the properties of the model studied here are analogous to 
those
of the usual $SU(2)$ Higgs model with symmetry breaking potential
supporting monopoles~\cite{mono}. Like in
that case, the Bogomol'nyi bound cannot be saturated, and, 
like-charged
solitons repel in the sense that a(n axially symmetric) multisoliton 
of charge
$n$ has higher energy than $n$ infinitely separated $1$-solitons. (It 
is
possible that solutions with less than axial symmetry may have lower 
mass than
the axially symmetric ones studied here, so the possibility exists 
that such
solutions may form bound states, however unlikely.) Another point in 
this
analogy is that the sphaleron describing soliton-antisoliton pairs 
here is
lighter than two $1$-solitons, just as it is in the Higgs 
case~\cite{KK}.


\bigskip
\noindent
{\bf Acknowledgement}
\\
We are grateful to Dieter Maison for useful remarks.
This work was carried out in the framework of Enterprise--Ireland
Basic Science Research Project SC/2003/390.

\end{document}